\documentclass{appolb}

\usepackage{epsfig}
\usepackage{graphicx}
\begin{document}

\title{Some properties of two Nambu--Jona-Lasinio -type models with inputs from lattice QCD
\thanks{Presented at eQCD2012}%
}

\author{P. Costa, C. A. de Sousa, M. C. Ruivo, O. Oliveira, P. J. Silva
\address{Centro de F\'isica Computacional, Departamento de F\'isica,
Universidade de Coimbra, P-3004-516 Coimbra, Portugal}
\and
{H. Hansen}
\address{IPNL, Universit\'e de Lyon/Universit\'e Lyon 1, CNRS/IN2P3, 4 rue E.Fermi, F-69622
Villeurbanne Cedex, France}
}
\maketitle


\begin{abstract}

We investigate the phase diagram of the so-called Polyakov--Nambu--Jona-Lasinio (PNJL) model at 
finite temperature and nonzero chemical potential. 
The calculations are performed in the light and strange quark sectors ($u$, $d$, $s$), 
which includes the 't Hooft instanton induced interaction term  that breaks the axial symmetry, 
and the quarks are coupled to the (spatially constant) temporal background gauge field. 
On one hand, a special attention is payed to the critical end point (CEP). 
The strength of the flavor-mixing interaction alters the CEP location, since when it becomes 
weaker the CEP moves to low temperatures and can even disappear.
On the other hand, we also explore the connection between QCD, a nonlocal Nambu--Jona-Lasinio 
type model and the Landau gauge gluon propagator. Possible links between the quenched gluon 
propagator and low energy hadronic phenomenology are investigated.

\end{abstract}

\section{The phase diagram in the context of the PNJL model }

Chiral symmetry breaking and confinement are two of the most important features of 
quantum chromodynamics (QCD). Chiral models like the Polyakov--Nambu--Jona-Lasinio (PNJL)
model have been successful in explaining the dynamics of spontaneous breaking of chiral 
symmetry and its restoration at high temperatures and densities/chemical potentials. 
The PNJL model also plays an interesting role in the investigation of the QCD phase structure.
Understanding the properties of matter at finite temperatures and densities is one of the 
most important goals from both the theoretical and experimental point of view. 
For example, the critical end point of QCD, proposed at the end of the eighties, 
is still a very important subject of discussion nowadays: indeed its existence and 
location is one of the main goals in SPS at CERN and in RHIC at BNL \cite{QCD_exp}.

The NJL model describes interactions between constituent quarks, giving the correct chiral properties;
static gluonic degrees of freedom are then introduced in the NJL lagrangian, through an effective gluon potential
in terms of Polyakov loops, with the aim of taking into account features of both chiral symmetry breaking
and deconfinement. The coupling of the quarks to the Polyakov loop leads to the reduction of the weight of 
quark degrees of freedom as the critical temperature is approached from above, which is interpreted as a 
manifestation of confinement and is essential to reproduce lattice results.

Our calculations are performed in the framework of an extended SU(3)$_f$ PNJL
Lagrangian, which includes the 't Hooft instanton induced interaction term that breaks
the U$_A$(1) symmetry, and the quarks are coupled to the (spatially constant) temporal
background gauge field $\Phi$ \cite{Fukushima,Ratti}:
%
\begin{eqnarray}\label{eq:lag} {\mathcal L}&=& \bar q(i \gamma^\mu D_\mu-\hat m)q +
\frac{1}{2}\,g_S\,\,\sum_{a=0}^8\, [\,{(\,\bar q\,\lambda^a\, q\,)}^2\, +
{(\,\bar q \,i\,\gamma_5\,\lambda^a\, q\,)}^2\,] \nonumber\\
&+& g_D\,\left\{\mbox{det}\,[\bar q\,(1+\gamma_5)\,q] +\mbox{det} \,[\bar q\,(1-\gamma_5)\,q]\right\}
- \mathcal{U}\left(\Phi[A],\bar\Phi[A];T\right).
\end{eqnarray}

The covariant derivative is defined as $D^{\mu}=\partial^\mu-i A^\mu$, with
$A^\mu=\delta^{\mu}_{0}A_0$ (Polyakov gauge); in Euclidean notation $A_0 = -iA_4$.  The
strong coupling constant $g$ is absorbed in the definition of $A^\mu(x) = g {\cal
A}^\mu_a(x)\frac{\lambda_a}{2}$, where ${\cal A}^\mu_a$ is the (SU(3)$_c$) gauge field
and $\lambda_a$ are the (color) Gell-Mann matrices.

The  effective potential for the (complex) field $\Phi$ adopted in our pa\-ra\-me\-tri\-za\-tion of
the PNJL model  reads:

\begin{eqnarray}
    \frac{\mathcal{U}\left(\Phi,\bar\Phi;T\right)}{T^4}
    =-\frac{a\left(T\right)}{2}\bar\Phi \Phi +
    b(T)\mbox{ln}\left[1-6\bar\Phi \Phi  + 4(\bar\Phi^3+ \Phi^3)-3(\bar\Phi
    \Phi)^2\right],
    \label{Ueff}
\end{eqnarray}
where
\begin{equation}
    a\left(T\right)=a_0+a_1\left(\frac{T_0}{T}\right)+a_2\left(\frac{T_0}{T}
  \right)^2\,\mbox{ and }\,\,b(T)=b_3\left(\frac{T_0}{T}\right)^3.
\end{equation}

The parameters of the effective potential $\mathcal{U}$ are given by $a_0=3.51$, $a_1=
-2.47$, $a_2=15.2$ and $b_3=-1.75$. 
When quarks are added, the parameter $T_0$, the critical temperature for the deconfinement 
phase transition (that manifests itself as a breaking of the center symmetry)
within a pure gauge approach, was fixed to $270$ MeV, according to lattice findings.
This choice ensures an almost exact coincidence between chiral crossover and deconfinement at
zero chemical potential, as observed in lattice calculations.

The parameters of the NJL sector are: $m_u = m_d = 5.5$~MeV,
$m_s = 140.7$ MeV, $g_S\Lambda^2 = 3.67$, $g_D \Lambda^5 = -12.36$
and $\Lambda = 602.3$ MeV, which are fixed to
reproduce the values of the coupling constant of the pion, $f_\pi\,=\,92.4$ MeV, and the
masses of the pion, the kaon, the $\eta$ and $\eta^\prime$, respectively,
$M_\pi\,=\,135$ MeV, $M_K\,=\,497.7$ MeV, $M_\eta\,=\,514.8$ MeV and
$M_{\eta^\prime}\,=\,960.8$ MeV \cite{varios}.

The inclusion of the Polyakov loop effective potential ${\cal U}(\Phi,\bar\Phi;T)$, that
can be seen as an effective pressure term mimicking the gluonic degrees of freedom of
QCD, is required to get the correct Stefan–-Boltzmann limit. 
Indeed in the NJL model the ideal gas limit is far to be reached due to the lack of 
gluonic degrees of freedom.

\begin{figure}[t]
\begin{center}
  \begin{tabular}{cc}
		\hspace*{-0.5cm}\includegraphics[width=0.575\textwidth]{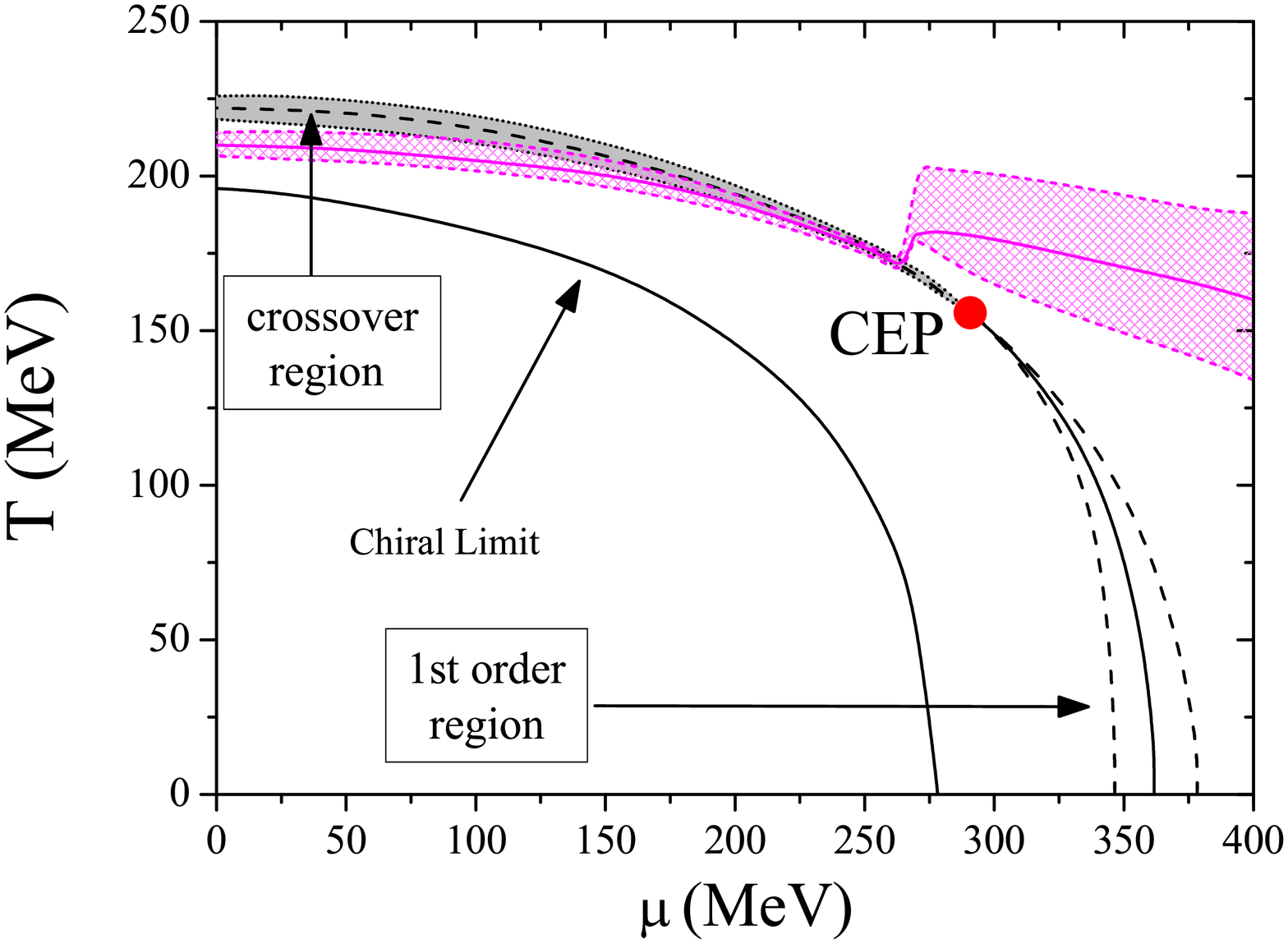}&
		\hspace*{-1.5cm}\includegraphics[width=0.575\textwidth]{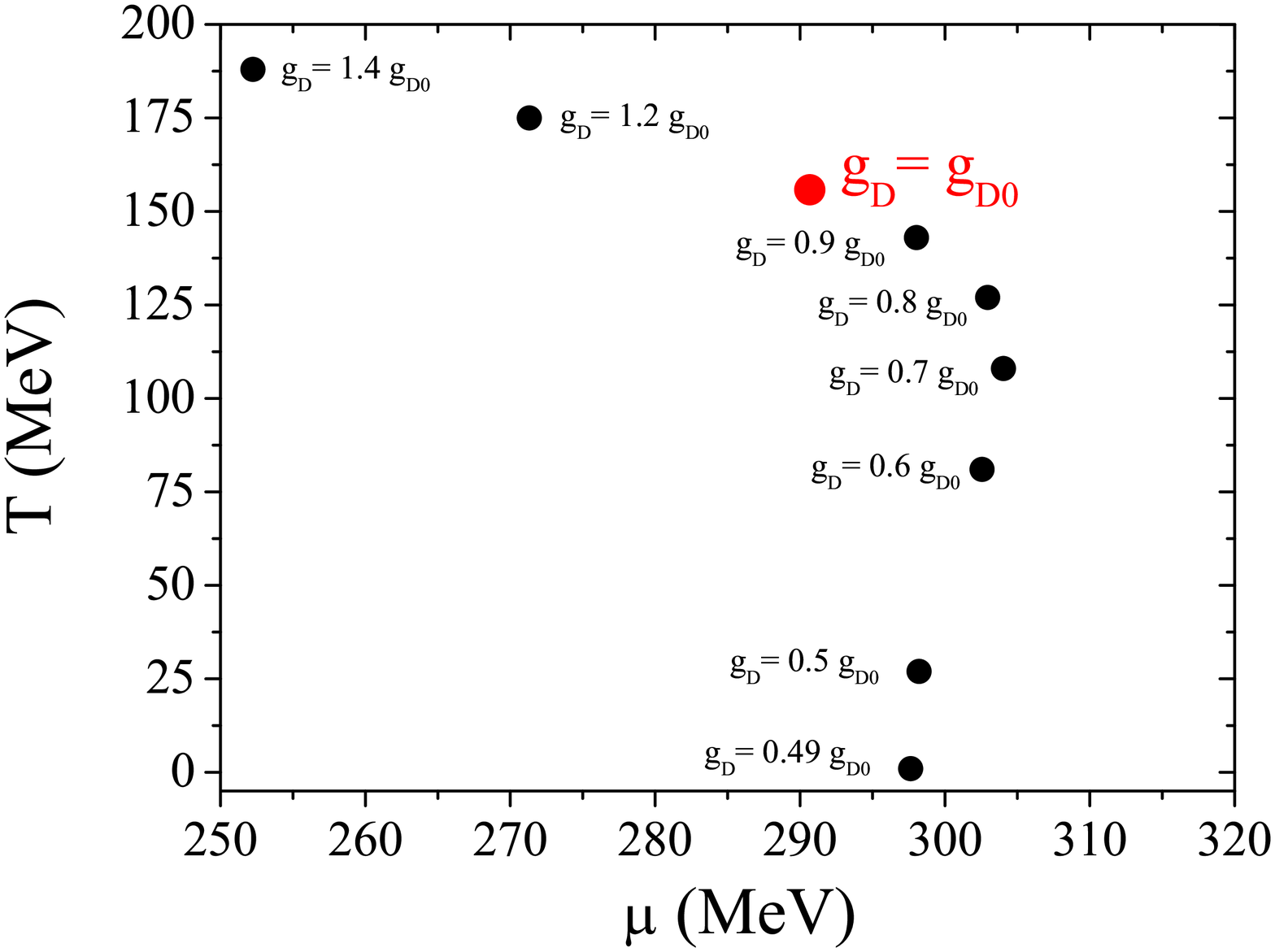}\\
	\end{tabular}
\end{center}
\caption{Phase diagram in the SU(3) PNJL model: the location 
of the CEP is found at $T^{CEP}=155.80$ MeV and $\mu^{CEP} = 290.67$ MeV 
(see details in text). Dependence of the location of the CEP on the strength of the 't Hooft 
coupling constant $g_D$.}
\label{Fig:Phase_Diagram}
\end{figure}

In Fig. \ref{Fig:Phase_Diagram} (left panel), we present the phase diagram of the PNJL model.
As the temperature increases the chiral transition is first order and persists up to the CEP. 
At the CEP the chiral transition becomes a second order one.
The location of the CEP is found at $T^{CEP}=155.80$ MeV and $\mu^{CEP} = 290.67$ MeV
($\rho_B^{CEP}=1.87\rho_0$).
For temperatures above the CEP there is a crossover whose location is calculated making use of  
$\partial^2\left\langle \bar{q}q\right\rangle/\partial T^2=0$,
\textit{i.e.} the inflection point of the quark condensate $\left\langle \bar{q}q\right\rangle$. 

The transition to the deconfinement is given by $\partial^2\Phi/\partial T^2=0$,
and is represented by the magenta line. The surrounding shaded area  that limits
the region where the crossover takes place is determined as the inflection point 
of the susceptibility $\partial\Phi/\partial T$.   

Due to the importance of the location of the CEP from the experimental point of view,
let us investigate the influence of other parameters which can lead to a significant 
change in the CEP's localization. 

It is well known that the U$_A$(1) anomaly has big influence on the behavior of 
several observables, so it is demanding to investigate possible changes in the 
location of the CEP in the $(T,\,\mu)$ plane when the anomaly strength is modified. 
The axial U$_A$(1) symmetry is broken explicitly by instantons,
leaving a SU(N$_f)\otimes$ SU(N$_f)$ symmetry which determines the
chiral dynamics. Since instantons are screened in a hot or dense
environment, the U$_A$(1) symmetry may be effectively restored in
matter. So, the change of the U$_A$(1) anomaly strength has a strong 
influence on the localization of the CEP in the $(T,\,\mu)$ plane. 

In Fig. \ref{Fig:Phase_Diagram} (right panel), we show the location
of the CEP for several values of $g_D$ compared to the results for $g_{D_0}$, 
the value used for the vacuum. 
As already pointed out by K. Fukushima in \cite{Fukushima}, 
the location of the CEP depends on the value of $g_D$. 
The results show that, in the framework of this model, the existence
or not of the CEP is determined by the strength of the anomaly
coupling, the CEP getting closer to the $\mu$ axis as $g_D$
decreases.
As the strength of the flavor-mixing interaction becomes weaker, the CEP moves 
to low temperatures and can even disappear.

\section{Low energy physics and the gluon propagator}

In this section we explore the connection between QCD, a nonlocal 
Nambu--Jona-Lasinio type model and the Landau gauge gluon propagator \cite{Costa:2010pp}. 

The interaction between quarks and gluons in QCD reads:
\begin{equation}
  \mathcal{L}_{\overline\psi\psi A} ~ = ~ g \, \overline\psi \, \gamma^\mu \, A^a_\mu \, \frac{\lambda^a}{2} \, \psi
\end{equation}  
Expanding the term containing $\mathcal{L}_{\overline\psi\psi A}$ up to 
$g^2$ and integrating the gluon fields (see \cite{Costa:2010pp} for details), 
the theory becomes an effective nonlocal fermionic theory
\begin{eqnarray}
S[\overline\psi,\psi]  &=&  \!\!\!\int d^4x d^4y \Big\{  \overline\psi (y) \, \delta (y-x) \, \left( 
           i \gamma^\mu \partial_\mu - m \right) \psi (x) \nonumber \\
  &+& \!\!\frac{g^2}{8}  \, J (x,y) D (x -y)   J (y,x)         -  
                \frac{g^2}{8} \, J_5 (x,y)  D (x -y)  J_5 (y,x) \Big\} 
   \label{EffAction}
\end{eqnarray}
with $J (x,y) =  \overline \psi (x) \psi (y)$ and $J_5 (x,y)  =  \overline \psi (x) \gamma_5 \psi (y)$
and $D (x,y)$ is the gluon propagator form factor.

First principles calculations of the gluon propagator have been performed using lattice QCD and DSE
(see for example \cite{qcdtnt11} and references therein). 
The momentum space propagator 
\vspace{-0.25cm}
\begin{equation}
        D(p^2)\,  = \, Z \, \frac{ \left(  p^2 \right)^{2 \kappa - 1}}{\left( p^2 + \Lambda^2_{QCD} \right)^{2 \kappa}} 
      \label{GlueProp}
\end{equation}
is able to describe both the scaling ($\kappa>0.5$) and decoupling ($\kappa=0.5$) 
infrared DSE solutions and the lattice data up to 
$p\sim800$ MeV; $\Lambda_{QCD}$ stands for an infrared mass scale.

\begin{figure}[t]
\begin{center}
  \begin{tabular}{cc}
		\hspace*{-0.5cm}\includegraphics[width=0.575\textwidth]{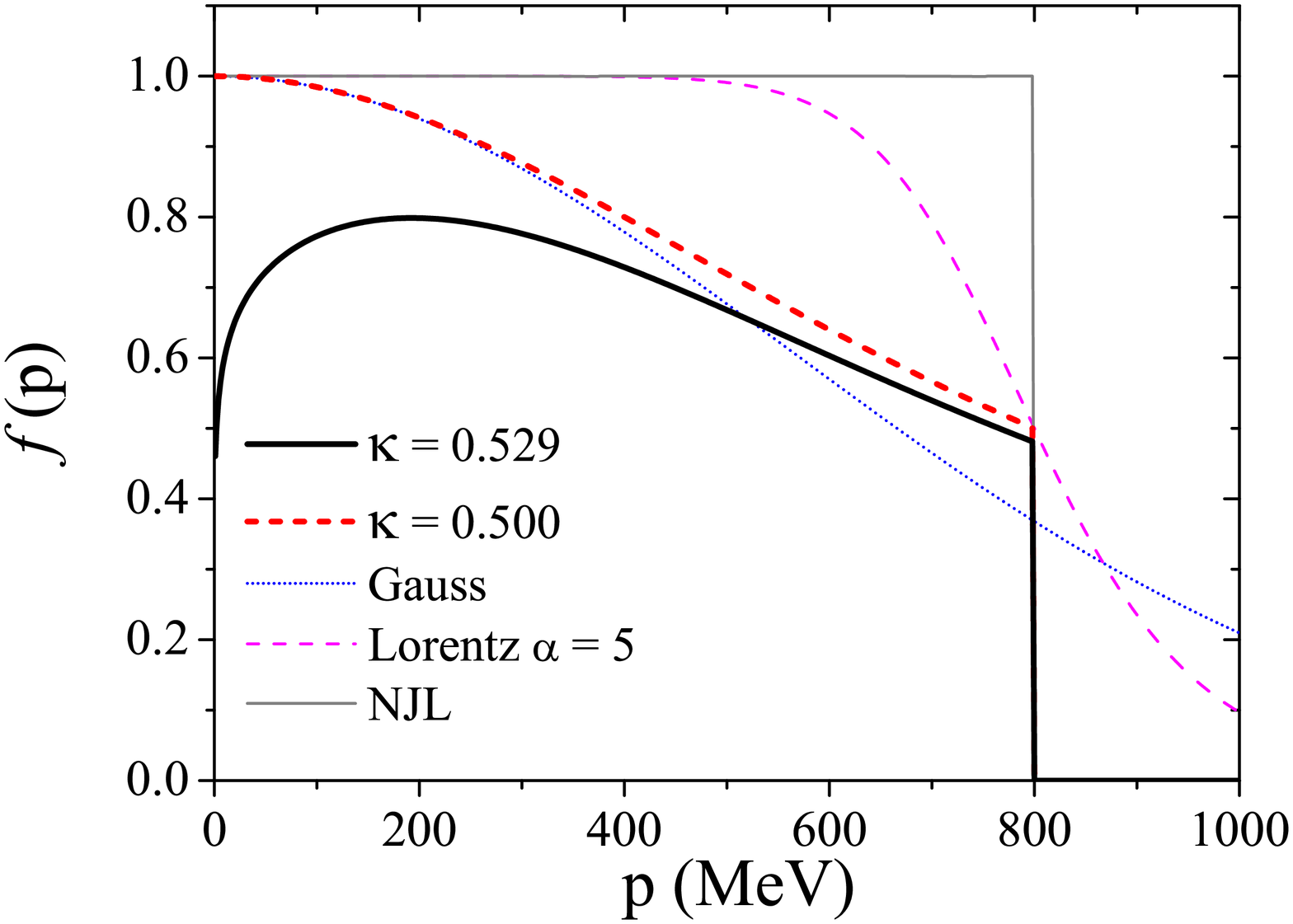}&
		\hspace*{-1.25cm}\includegraphics[width=0.575\textwidth]{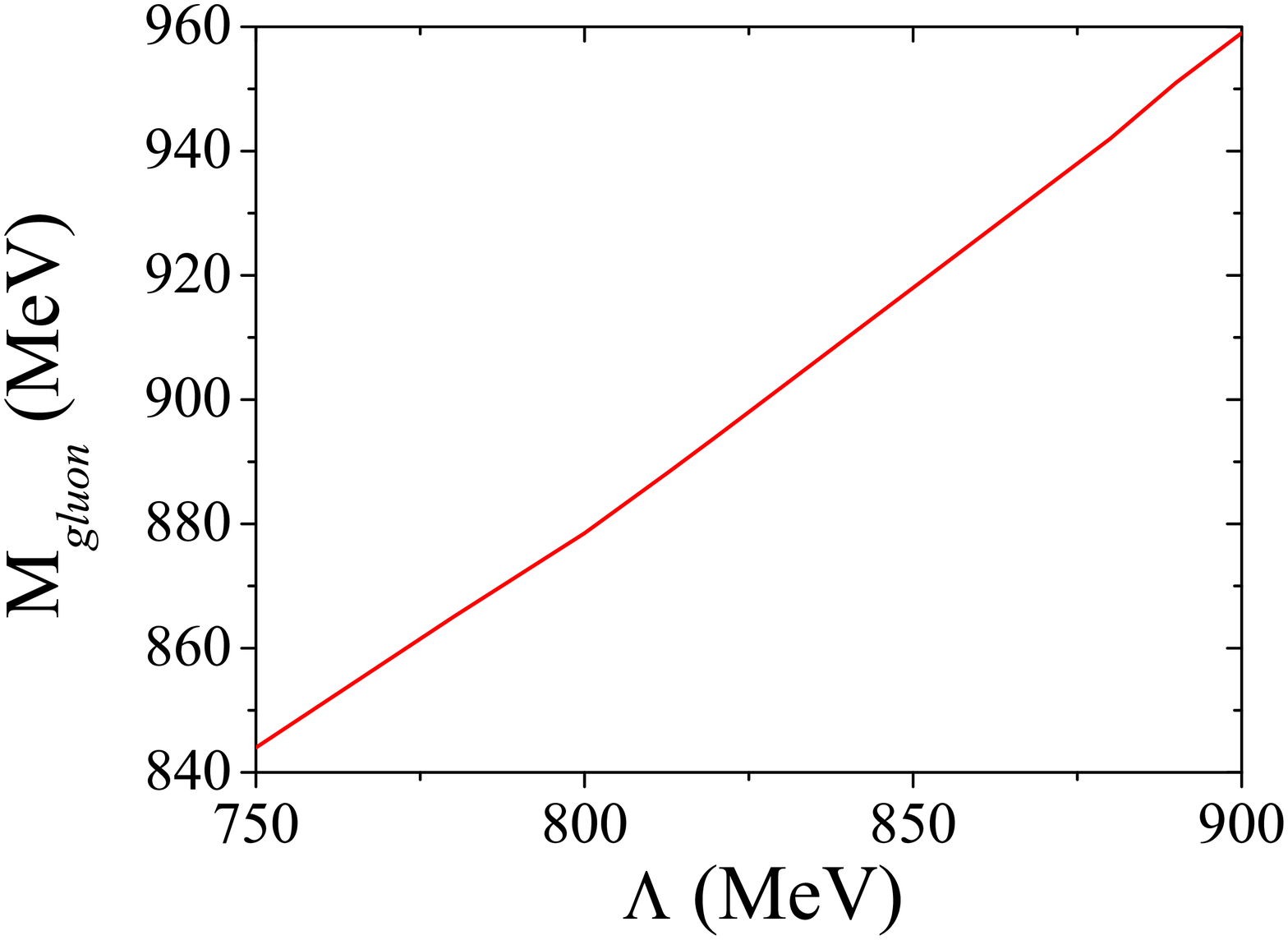}\\
	\end{tabular}
\end{center}
\label{fig:dec2}
\caption{Form factors as a function of $\mbox{p}$. The figure includes typical form factors used in  previous studies.
$M_{gluon} (\Lambda)$ required to reproduce the experimental $\Gamma_{\pi\rightarrow\gamma\gamma}$.}
\end{figure}

Let us define the dimensionless form factor in momentum space as
\begin{equation}
f({p^2}) ~ = ~ \Lambda^2 D(p^2) ~ = ~ \frac{\Lambda^2}{{p}^2} \, 
         \left(\frac{{p}^2}{{p}^2 + \Lambda^2_{QCD}}\right)^{2\kappa} \theta(\Lambda-{p}) \, .
\label{form_factorS}
\end{equation}
The constant $Z$  in Eq. (\ref{GlueProp}) will be included in the 
definition of the coupling constant $G$, which multiplies the quark 
currents of the nonlocal theory.  $G$ carries the dimension of a length
squared.  In $f(p^2)$, $\Lambda$ is the cut-off. In a first step
we assume $\Lambda_{QCD} = \Lambda$.
The form factor $f(p^2)$ is shown in Fig. \ref{fig:dec2} (left panel), 
together with typical form factors considered in the literature. 

Demanding the nonlocal model to reproduce the experimental values for  
$M_\pi$, $f_\pi$ and $\Gamma_{\pi \rightarrow \gamma\gamma}$, with a cut-off 
$\Lambda=\Lambda_{QCD}=800$ MeV, we obtain $m_q =4.205$ MeV, 
$-\left\langle \bar{q}q \right\rangle^{1/3}= 271.1$ MeV, $G\Lambda^2 = 7.491$
and $\kappa = 0.529 $.
The presented results favor $\kappa > 0.5$. Now we investigate the decoupling type of 
propagator 
\begin{equation}
D({p^2}) ~ = ~ \frac{Z}{{p}^2 + M^2_{gluon}},
\end{equation}
where $M_{gluon}$ takes the role of $\Lambda_{QCD}$ and can be interpreted as an effective gluon mass.
Requiring the model to reproduce the same experimental quantities as before, 
we find $M_{gluon}$ within typical values found in the literature, but with a strong 
dependence with the cut-off. 
In fact, $M_{gluon}$ is a linear function of $\Lambda$
-- see Fig. \ref{fig:dec2}-- right panel. In what concerns the quark condensate, 
the model shows that $\langle \overline q q \rangle$
increases with $M_{gluon}$, i.e with the cut-off $\Lambda$. 
In order to reproduce the experimental value of the 
condensate, i.e. to have $(- \left\langle \bar{q}q \right\rangle)^{1/3} = 270$ 
MeV, it turns out that the gluon mass is
$M_{gluon} = 878$ MeV for $\Lambda = 800$ MeV.
We therefore conclude that low energy physics does not distinguish between the so-called 
decoupling and scaling solutions of the Dyson-Schwinger equations. 
This result means that, provided that the model parameters are chosen appropriately, 
one is free to choose any of the above scenarios. 

Finally, it is interesting to refer that the model considered here is chiral invariant 
and satisfies the GMOR relation at the 1\% level of precision.

As future work we will generalize the results to non-zero temperature (this requires
modeling the gluon propagator at finite temperature by a functional form compatible with
both Dyson-Schwinger and lattice QCD results) which will allow us to 
investigate the meson properties at finite temperature as probes for the
chiral symmetry restoration.

\vskip0.25cm

Work supported by projects CERN/FP/116356/2010 and PTDC/FIS/ 100968/2008, 
projects developed under the initiative QREN financed by the UE/FEDER through
the Programme COMPETE - ``Programa Operacional Factores de Competitividade''.



\vspace{-0.4cm}
\begin{thebibliography}{9}

\bibitem{QCD_exp}
M.M. Aggarwal  et al. (STAR Collaboration),
arXiv:1007.2613 [nucl-ex];
Terence J. Tarnowsky,
J. Phys. G38 (2011) 124054.

\bibitem{Fukushima}
K. Fukushima,
Phys. Rev. D77 (2008) 114028.

\bibitem{Ratti}
C. Ratti, M. A. Thaler, W. Weise,
Phys. Rev. D73 (2006) 014019.

\bibitem{varios}
P. Costa, M. C. Ruivo, C. A. de Sousa, H. Hansen, and W.M. Alberico,
Phys. Rev. D79 (2009) 116003;
P. Costa, C. A. de Sousa, M. C. Ruivo, and H. Hansen,
Europhys. Lett. 86 (2009) 31001;
P. Costa, H. Hansen, M. C. Ruivo, and C. A. de Sousa,
Phys. Rev. D81 (2010) 016007;
P. Costa, M. C. Ruivo, C. A. de Sousa, and H. Hansen,
Symmetry 2 (2010) 1338.

\bibitem{Costa:2010pp} 
P.~Costa, O.~Oliveira and P.~J.~Silva,
Phys.\ Lett.\ B {695} (2011) 454.

\bibitem{qcdtnt11}
O.~Oliveira, P.~J.~Silva and P.~Bicudo,
PoS FACESQCD {\bf } (2010) 009.

\end{thebibliography}
\end{document}